\documentclass[twocolumn,showpacs,preprintnumbers,amsfonts]{revtex4}

\usepackage{bm}

\begin{document}


\title{Fine-structure constant variability: surprises
for  \\ laboratory atomic spectroscopy and cosmological
evolution of quasar spectra }

\author{Jacob D.
Bekenstein}\email{bekenste@vms.huji.ac.il}
\homepage{http://www.phys.huji.ac.il/~bekenste/}
\affiliation{Racah Institute of Physics, Hebrew
University of Jerusalem\\ Givat Ram, Jerusalem 91904
ISRAEL}

\date{\today}

\begin{abstract} Calculation of the Dirac hydrogen
atom spectrum in the framework of dynamical fine
structure  constant ($\alpha$) variability discloses a
small  departure in the laboratory from Sommerfeld's 
formula for the fine structure shifts, possibly measurable
today.  And for a distant object in the universe, the
wavelength shift of a spectral line specifically
ascribable to cosmological $\alpha$ variation is
found to depend differently on the quantum numbers
than in the conventional view.  This last  result clashes
with the conventional wisdom that an
atom's spectrum can change with cosmological time only
through evolution of the $\alpha$ parameter in the
energy eigenvalue formula, and thus impacts on the
Webb group's analysis of fine structure intervals in
quasar absorption lines (which has been claimed to
disclose  cosmological $\alpha$ evolution). In
particular, analyzing together a mix of quasar
absorption lines from different fine structure
multiplets can bias estimates of cosmological
$\alpha$ variability. 
\end{abstract}


\maketitle 

Early theoretical speculations about temporal
variability of the fine-structure constant $\alpha$
\cite{jordan,teller-gamow,dicke} (an excellent review
is provided by Uzan~\cite{uzan}) have assumed urgency
in view of the claimed cosmological variability in
fine-structure multiplets seen in absorption in
distant quasar spectra~\cite{webb1,webb2}.    In the
Webb group's analysis it is assumed that the spectrum
of an atom in an intergalactic cloud at redshift $z$
can be gotten from its laboratory spectrum by merely
replacing the laboratory value of $\alpha$ by that at
redshift $z$.  This assumption is crucial to the
group's strategy  for reducing experimental errors by
analyzing simultaneously  lines from different
fine-structure multiplets of different  elements.

The mentioned assumption takes it for granted that in
a world where $\alpha$ varies, the \textit{form} of
the Dirac hamiltonian (the most direct tool for
determining the fine structure splittings) relevant
for an electron in an atom is  the same as in standard
physics.  Is such an ``electromagnetic equivalence
principle''  correct?

In one paradigm cosmological $\alpha$
variability~\cite{grojean} occurs in discrete jumps,
each precipitated by a  phase transition in which
spontaneous  symmetry breaking changes a certain
vacuum expectation value, and thereby the value of
$\alpha$.  In this point of  view it seems reasonable
for the  Dirac Hamiltonian to retain  its textbook
form, at least during each cosmological interval  of
constant $\alpha$.

Things are different in the much earlier paradigm,
first contemplated by Jordan~\cite{jordan} and
Dicke~\cite{dicke}, where $\alpha$ is a dynamical
field.  Both workers emphasized that in a covariant
theory, temporal dynamical $\alpha$ variability must
be accompanied by spatial variability.  Concerns,
first  articulated by Dicke~\cite{dicke}, that
spatial variability necessarily entails observable
violations of the weak  equivalence principle have
proved groundless, at least within  the general
framework of dynamical $\alpha$ variability  proposed
by us~\cite{bek}: its violations of the equivalence 
principle lie beyond foreseeable experimental
sensitivities~\cite{bek2}.

As we show here for the hydrogen atom, dynamical
$\alpha$ variability according to the framework
modifies the form of the Dirac hamiltonian. One
consequence is a small deviation of the hydrogen
atom's spectrum in today's laboratory from that given
by Sommerfeld's formula [Eq.~(\ref{spectrum}) below].
Dynamical $\alpha$ variability also implies
nontrivial evolution of atomic spectra with 
cosmological epoch (time). Estimation of $\alpha$'s
cosmological evolution from quasar fine-structure 
multiplets of mixed provenance must take cognizance of
this effect. 

In its most convenient form, the framework for
$\alpha$ variability is a theory of \textit{constant}
charges where the permittivity and the reciprocal
permeability of the vacuum both equal $e^{-2\psi}$,
where $\psi$ is a scalar field~\cite{bek2,SBM}.  One
consequence is that the coupling ``constant''
$\alpha$ of Coulomb's law, which is inversely
proportional to the vacuum's permittivity,  varies as
$e^{2\psi}$. Another is that the electromagnetic
field tensor
$f^{\mu\nu}$ and its dual ${}^*\! f^{\mu\nu}$ obey
the equations
\begin{eqnarray} {}^*\! f^{\mu\nu}{}_{;\nu}&=&0,
\label{dual}
\\ (e^{-2\psi}f^{\mu\nu})_{;\nu}&=&{4\pi\over c} j^\mu
\label{Maxwell}
\end{eqnarray} with $j^\mu$ the electric current. 
The coupling between
$e^{-2\psi}$ and $f^{\mu\nu}$ induces an
electromagnetic contribution to the source term of
the scalar field equation,
\begin{equation}
\psi_{,\mu;}{}^\mu =4\pi\kappa^2\sigma-{\kappa^2\over
2}e^{-2\psi}f_{\mu\nu}f^{\mu\nu},
\label{wave_equation}
\end{equation} where the density $\sigma$ is
associated with particles and
$\kappa$ is the parameter of the theory with
dimensions of reciprocal energy times charge,.  
Whenever
$j^\mu$ can be described as a sum of contributions
from discrete particles,
$\sigma$ is obtained by replacing in $j^\mu$  the
product of each particle's charge and the
corresponding 4-velocity by the derivative
$\partial m c^2/\partial \psi$ of the particle's mass
$m$.  So in this theory particle mass is, in
principle, a function of the ambient $\psi$ field. 

In flat spacetime the \textit{physical} static
solution of Eqs.~(\ref{dual})-(\ref{wave_equation})
for a pointlike charge
$e_0$ in terms of its formal Coulomb potential
$\Phi=e_0/r$ and its true electric field,
$\bm{E}\equiv c\{f^{01},f^{02},f^{03}\}$,
is~\cite{bek2}
\begin{eqnarray}
\bm{E}&=&-e^{2\psi}\bm{\nabla}
\Phi=-\bm{\nabla}(\kappa^{-1}\tan(\kappa\Phi+\chi))
\label{E}
\\ e^\psi&=&\sec(\kappa\Phi+\chi),
\label{epsi}
\end{eqnarray} with $\chi$ a constant of integration.
A second such constant has been set to unity by
redefinition of  the charge~\cite{bek2}.  This
restricts our description to epochs for which
$e^\psi>1$. Now $\chi$ is evidently equivalent to 
$\exp(\psi_{\rm cosm})$, the \textrm{cosmological}
value of $e^{\psi}$ at the relevant epoch, which plays
the role of asymptotic value of $e^{\psi(\bm{r})}$. 
Solution (\ref{E})-(\ref{epsi})  thus depends on an
\textit{adiabatically} varying parameter.  We assume that
$|\chi|\ll 1$ throughout the epochs of interest.

Solution (\ref{E})-(\ref{epsi}) applies as well to
multiple pointlike particles at rest~\cite{bek2} with
$\Phi$ replaced by the Coulomb potential coming from
all the charges. According to Ref.~\onlinecite{bek2},
such a  solution is consistent arbitrarily near a
source only if the latter's mass
$m$ is subject to the condition 
$\kappa^2\partial m c^2/\partial \psi =
-e_0d\psi/d\Phi$ with both sides \textit{evaluated
at} the relevant particle's position.  The theory
thus fixes each $m$'s dependence on
$\psi$.  By Eq.~(\ref{epsi})  the condition is
\begin{equation}
\partial m c^2/\partial \Phi =
-e_0\tan^2(\kappa\Phi+\chi)
\label{change}
\end{equation}  The derivative here gives the change
of $m$ reflecting a local change of the total Coulomb
potential $\Phi$ caused by charges elsewhere (the
charge's self-potential,
$\Phi_\textrm{s}$, cannot change, of course; since
the framework does not allow truly point charges,
$\Phi_\textrm{s}<\infty$~\cite{bek2}). 

The integral of Eq.~(\ref{change})  is
\begin{equation}
mc^2=C-e_0\kappa^{-1}\tan(\kappa\Phi+\chi)+e_0\Phi
\label{integral}
\end{equation}   with $C$ an integration constant. 
In terms of the free particle's rest mass $m_0$
($m\rightarrow m_0$ for
$\Phi\rightarrow\Phi_\textrm{s}$) at the epoch in
question, 
$C=m_0+e_0\kappa^{-1}\tan(\kappa\Phi_\textrm{s}+\chi)
-e_0\Phi_\textrm{s}$.  Now from
Ref.~\onlinecite{bek2} it follows that $\kappa^{-1}$
is no more than an  order of magnitude below Planck's
energy divided by the elementary charge.   Thus, for
example, even at distance $10^{-17}$ cm  from an
electron (if such accuracy in position has any
meaning in light of the
\textit{Zitterbewegung}), $e_0\Phi$ receives a
contribution 18 orders \textit{below} Planck energy
and $\kappa|\Phi|$ and $\kappa|\Phi_\textrm{s}|$ are 
both smaller than $10^{-18}$.   Since we have no 
reasons to expect that $\chi$ is this tiny at a 
generic epoch, we may regard both
$\kappa|\Phi|$ and $\kappa|\Phi_\textrm{s}|$ as small
compared to
$\chi$.   Thus for a (pointlike) electron with charge
$e_0$ immersed  in the Coulomb potential 
$\Phi_\textrm{n}$  of a nucleus at a cosmological
epoch specified by $\chi$, we obtain by expanding 
Eq.~(\ref{integral})  to first order in
$\kappa\Phi_\textrm{n}=\kappa(\Phi-\Phi_\textrm{s})$,
and   neglecting in the tangent's argument
$\kappa\Phi_\textrm{s}$  as compared to $\chi$ that
\begin{equation} m=m_0+\delta m\approx
m_0-e_0c^{-2}\tan^2\chi\cdot
\Phi_\textrm{n}.
\label{dm}
\end{equation}

Consider now the Dirac hamiltonian
$\hat H$ for the electron in an electric field
$\bm{E}$.  In an inertial frame
\begin{equation}
\hat H=(-\imath \hbar
c\bm{\alpha}\cdot\bm{\nabla}+(m_0+\delta m)c^2\beta +
e_0U I)
\label{dirac_ham}
\end{equation}  with $I$ the $4\times 4$ unit matrix
and
 $\beta$ and the triplet $\bm{\alpha}$ the four
familiar Dirac matrices~\cite{merz}.  By minimal
coupling the potential $U$ in which the electron
moves must be such that
$\bm{E}=-\bm{\nabla} U$.  From Eq.~(\ref{E}) we infer
that
\begin{equation} 
U=\kappa^{-1}\tan(\kappa\Phi+\chi)
-\kappa^{-1}\tan(\kappa\Phi_\textrm{s} +\chi ),
\end{equation}  
at the electron's position, the added
constant ensuring that $U$ vanishes asymptotically
for any $\chi$, just as does
$\Phi_\textrm{n}$.  We include the electron's
$\Phi_\textrm{s}$ in $\Phi$ because of nonlinearity
of the theory [see Eqs.~(\ref{E})-(\ref{epsi})].
Expanding $U$ to
$\mathcal{O}(\Phi_\textrm{n})$ and again dropping
$\kappa\Phi_\textrm{s}$ in comparison with $\chi$, we
get
\begin{equation} U\approx \sec^2\chi\cdot
\Phi_\textrm{n}
\label{tilde_Phi}.
\end{equation}

With help of Eqs.~(\ref{dm}) and (\ref{tilde_Phi}),
Eq.~(\ref{dirac_ham}) may be rewritten as
\begin{eqnarray}
\hat H&=&\hat H_0 + \delta \hat H
\label{total}
\\
\hat H_0&=&(-\imath \hbar
c\bm{\alpha}\cdot\bm{\nabla}+m_0c^2\beta +
e_0\Phi_\textrm{n}\ I)
\label{zeroth}
\\
\delta\hat
H&=&(I-\beta)e_0\tan^2\chi\cdot\Phi_\textrm{n}
\end{eqnarray} Now the mass of the nucleus (proton)
is also subject to the same ``correction'' as in
Eq.~(\ref{dm}): the proton's charge is $-e_0$ but it
feels the
\textit{electron}'s potential $-\Phi_\textrm{n}$. 
However, we do not include this mass ``correction'' in
$\hat H$ just as we do not add an extra
$(-e_0)(-\Phi_\textrm{n})$ on account of the nucleus
to the
\textit{potential} term in Dirac's equation.  We
likewise drop from $\hat H$ the free proton's mass
(analog of $m_0$)  since it  cannot affect transition
wavelengths.      

Eq.~(\ref{zeroth}) is the textbook Dirac  hamiltonian
for an electron (charge
$e_0$) in the external Coulomb potential
$\Phi_\textrm{n}$.   Let $\alpha_*\equiv
e_0{}^2/\hbar c$ (to be distinguished from the
coupling ``constant''
$\alpha$). In terms of the principal quantum number
$n=1,2,\cdots\,$ and  the total angular momentum
quantum number
$j={\scriptstyle 1\over \scriptstyle 2},
{\scriptstyle 3\over
\scriptstyle 2},\cdots\,$, its eigenvalues, to
$\mathcal{O}(\alpha_*{}^4)$, are~\cite{LL} 
\begin{equation}
E_0=m_0c^2\left[1- {\alpha_*{}^2\over
2n^2}-{\alpha_*{}^4\over 2n^3}\left({1\over
j+{\scriptstyle 1\over
\scriptstyle 2}}-{3\over 4n}
\right)\right].
\label{spectrum}
\end{equation}    The term of
$\mathcal{O}(\alpha_*{}^2)$ is the Bohr energy level;
that of
$\mathcal{O}(\alpha_*{}^4)$ is Sommerfeld's  fine
structure correction.

In Eq.~(\ref{total}) $\delta \hat H$  represents the
perturbation coming from cosmological
$\alpha$ evolution.  We shall now show that it
modifies the conventional expression for the
wavelength
$\lambda$ of the transition
$n\rightarrow n-\Delta n$, $j\rightarrow j-\Delta j$
to
\begin{eqnarray} {1\over \lambda}&=&{m_0
c\alpha_*{}^2\over 4\pi\hbar}\times
\label{new}
\\
\Big[\Delta{1\over n^2}
&+&\alpha{}^2\Delta\Big({1\over n^3(j+{\scriptstyle
1\over \scriptstyle 2})}-{3\over
4n^4}\Big)+{\scriptstyle 1\over \scriptstyle
2}\alpha(\alpha-\alpha_*)\Delta{1\over n^4}\Big].
\nonumber
\end{eqnarray} This may be contrasted with the
conventional one--parameter formula  coming just from
Eq.~(\ref{spectrum}), namely Eq.~(\ref{new})  with
$\alpha_*\mapsto\alpha$ everywhere.

If $\Xi$ denotes the unit-normalized eigenspinor of
the eigenvalue problem
$\hat H_0\Xi=E_0\Xi$, then to first order in
perturbation theory the cosmological evolution shift
in the energy level is
\begin{eqnarray}
\delta E &=& \mathcal{I}\tan^2\chi
\label{shift}
\\
\quad \mathcal{I}&\equiv& \int
\Xi^\dagger (I-\beta)\Xi\ e_0\Phi_\textrm{n}\ d^3 x
\label{J}
\end{eqnarray} What is $\tan^2\chi$ in
Eq.~(\ref{shift}) ?  Since
$\alpha/\alpha_*$ is the reciprocal permittivity of
the vacuum, and $ \exp(2\psi_{\rm cosm})=\sec^2\chi$,
we may write
\begin{equation}
\tan^2\chi=\sec^2\chi-1=(\alpha-\alpha_*)/\alpha_*
\label{tan}
\end{equation}  Thus the permittivity is unity when
$\chi=0$.  The precise epoch in question depends on the
cosmological model; we defer consideration of this
issue to a future publication.  Our choice precludes
study of epochs with
$\alpha<\alpha_*$.

To calculate $\mathcal{I}$ one first notes that
$I-\beta$ is the matrix
$\textrm{diag}(0,0,2,2)$ so that with $\Xi$ written
as the bispinor
${\xi\choose \eta}$,
\begin{equation} \mathcal{I}=2\int \eta^\dagger
\eta\ e_0\Phi_\textrm{n}\ d^3 x.
\label{J1}
\end{equation} Now in the usual representation of
$\bm{\alpha}$ in terms of the Pauli matrices
$\bm{\sigma}$, 
$\hat H_0\Xi=E_0\Xi$ takes the
form~\cite{merz}
\begin{eqnarray} -\imath\hbar
c\bm{\sigma}\cdot\bm{\nabla}\eta+m_0c^2\xi+e_0\Phi_\textrm{n}\xi=
E_0\xi,
\label{xi}
\\ -\imath\hbar
c\bm{\sigma}\cdot\bm{\nabla}\xi-m_0c^2\eta+e_0\Phi_\textrm{n}\eta=
E_0\eta.
\label{eta}
\end{eqnarray} By requiring, in spherical polar
coordinates
$\{r,\theta,\varphi\}$, that
$\Xi$ be an eigenfunction of the $z$-component as
well as of the square of the total angular momentum,
$\hat\mathbf{J}$ (eigenvalues 
$m\hbar$ and $j(j+1)\hbar^2$, respectively), one is
led to the two possible forms~\cite{merz}
\begin{equation}
\Xi={F(r){\cal Y}^{jm}_{j-1/2}\choose -if(r){\cal
Y}^{jm}_{j+1/2}}
\quad\textrm{or}\quad
\Xi={G(r){\cal Y}^{jm}_{j+1/2}\choose -ig(r){\cal
Y}^{jm}_{j-1/2}}
\label{uu}
\end{equation}  with ${{\cal
Y}_\ell}^{jm}(\theta,\varphi)$ a spinorial spherical
harmonic.  The two forms differ in their parity for
given $j$; the first corresponds to the  case 
$\ell=j-{\scriptstyle 1\over \scriptstyle 2}$ in
nonrelativistic  theory of the hydrogen atom, while
the second has
$\ell=j+{\scriptstyle  1\over \scriptstyle 2}$.  If
the spinorial harmonics are normalized according to
$\int{{\cal Y}_\ell}^{jm}{}^{\dagger}\ {{\cal
Y}_\ell}^{jm}\,
\sin\theta\ d\theta d\varphi =1$, then the
normalization of
$\Xi$ requires
$\int_0^\infty (F^2+f^2)r^2 dr=1$  and similarly for
$G$ and
$g$ (as will be clear all four functions can be
chosen real).  

In terms of the dimensionless variables
$x=rm_0c/\hbar$,  
$\epsilon={\cal E}_0/m_0c^2$ and
$k=j+{\scriptstyle 1\over \scriptstyle 2}$, the
equations for the first form of $\Xi$  can  be cast
into the form~\cite{merz}
\begin{eqnarray}
(\epsilon-1+\alpha_*/x)F-[d/dx+(k+1)/x]f=0,
\\ (\epsilon+1-\alpha_*/x)f+[d/dx+(k-1)/x]F=0
\end{eqnarray} These admit the series solution
\begin{eqnarray} F=e^{\sqrt{1-\epsilon^2}x} x^\gamma
\sum_{\nu=0} a_\nu x^\nu,
\nonumber
\\ f=e^{\sqrt{1-\epsilon^2}x} x^\gamma \sum_{\nu=0}
b_\nu x^\nu.
\label{tt}
\end{eqnarray} Consistency between the two relations
connecting
$a_0$ and $b_0$ requires that 
\begin{equation}
\gamma=-1+\sqrt{k^2-\alpha_*{}^2}
\end{equation} (the negative  root is unphysical). 
In order for the sums in  Eqs.~(\ref{tt}) to
terminate at
$\nu= n'=0,1,
\cdots\ $ (otherwise $F$ and $f$ would diverge), one
must have
\begin{equation}
\epsilon=\left[1+{\alpha_*{}^2\over[\sqrt{k^2-\alpha_*{}^2}+n']^2}
\right]^{-1/2}.
\label{full}
\end{equation}  Expanding this to ${\cal
O}(\alpha_*{}^4)$ gives the Sommerfeld spectrum,
Eq.~(\ref{spectrum}), with
$n\equiv n'+j+{\scriptstyle 1\over \scriptstyle 2}$.  

Because $\Phi_n$ is spherically symmetric, the
normalization of the spinorial harmonics allows us to
write
\begin{equation} \mathcal{I}=2\int_0^\infty f^2\
e_0\Phi_\textrm{n}\ r^2 dr,
\end{equation} where $\Phi_n=-e_0/r$.   If $\Xi$ is
not normalized to unity, then in passing to the
integration variable
$x$ we must write
\begin{equation} \mathcal{I}=-{2m_0
c^2\alpha_*\int_0^\infty f^2 x dx\over
\int_0^\infty (F^2+f^2) x^2 dx }.
\label{ratio}
\end{equation}

The exact relativistic wavefunctions~\cite{LL} are
rather cumbersome, but are  not needed if all we want
is to calculate
$\mathcal{I}$ to
${\cal O}(\alpha_*{}^4)$.  Thus starting with
$a_0=1$, we used {\sl Mathematica\/} to  work out,
for a definite
$n'$, all the other $a_\nu$ and $b_\nu$ to ${\cal
O}(\alpha_*{}^{n'+1})$ by including in the recursion
relations~\cite{merz} expressions for 
$\gamma$ and  $\epsilon$ correct to
${\cal O}(\alpha_*{}^{n'+1})$.  The series in $F$ and
$f$ [see Eq.~(\ref{tt})] turn out to be of the form
$\sum_{i=0} U_{i}(\alpha_* x)
\alpha_*{}^{2i}$ and $\alpha\sum_{i=0} V_{i}(\alpha_*
x)
\alpha_*{}^{2i}$, where
$U_{i}$ and $V_{i}$ are polynomials of order $n'$. 
Up to normalization, $U_0(\alpha_* x)$ is the
associated Laguerre polynomial familiar from
nonrelativistic theory of the hydrogen  atom. The
correctness of the so obtained $F$ up to terms with
$i=1$ and of the $i=0$ term in $f$  was checked by
verifying that the Hellmann-Feynman
theorem~\cite{merz},
\begin{equation} {\partial E_0\over \partial
\alpha_*}=
\langle \Xi|{\partial \hat H_0\over
\partial\alpha_*}|\Xi
\rangle =  m_0 c^2{\int_0^\infty (F^2+f^2) x dx\over
\int_0^\infty (F^2+f^2) x^2 dx },
\end{equation} is satisfied with $E_0$
given by Eq.~(\ref{spectrum}).

A  change in the integration variable in
Eq.~(\ref{ratio}) to
$y=\alpha_* x$ (essentially the radial coordinate in
units  of the Bohr  radius) brings out an extra
factor $\alpha_*$ in the ratio of integrals.  Since
$f$ carries an overall factor of $\alpha_*$, the
$\mathcal{O}(\alpha_*{}^4)$ contribution
$\delta E$ can be computed by including only
the $i=0$ terms of $f$ in the numerator of
Eq.~(\ref{ratio}) and of
$F$ in the denominator, and totally  discarding $f$
in the denominator.  The result of exact integrations
is
\begin{equation}
\mathcal{I}=-{m_0c^2\alpha_*{}^4\over n^3}\left(
{1\over j+{\scriptstyle 1\over \scriptstyle
2}}-{1\over 2n}\right)
\label{I}
\end{equation}

The equations for the second form of $\Xi$ in
Eq.~(\ref{uu}) are obtained from
Eqs.~(\ref{tt})-(\ref{full})   by the replacements
$F\mapsto G$, $f\mapsto g$ and $k\mapsto -k$.  The
coefficients 
$a_\nu$ and $b_\nu$ are thereby changed, but $\gamma$
and
$\epsilon$ are not, except that now
$n'=1,2,\cdots $ ($n$ is still $n'+j+{\scriptstyle
1\over
\scriptstyle 2}$).  Repeating the procedure outlined
above we find the analog of $\mathcal{I}$ to have
exactly the same form as Eq.~(\ref{I}).  One
consequence is that, at least to
$\mathcal{O}(\alpha_*{}^4)$, the energies of levels
with like
$j$ but opposite parity are still degenerate, as in
the exact Dirac hydrogen spectrum. 

From Eqs.~(\ref{spectrum}), (\ref{shift}),
(\ref{tan}) and (\ref{I}) we deduce formula
(\ref{new}) for the transition wavelength.  It shows
an unconventional dependence on the cosmological
``fossil'' $\alpha_*$, the value of $\alpha$ when
the vacuum's permittivity was unity.  In principle
fitting a generic collection of laboratory measured
transition wavelengths with formula (\ref{new})
should allow separate determination of
$\alpha$ and $\alpha-\alpha_*$.  A reliable nonzero
result for
$\alpha-\alpha_*$ (which if the Webb group is right,
might be of order $10^{-5}\alpha$) would establish
dynamical
$\alpha$ variability according to the theories
represented by the framework as opposed to phase
transition variability.  Possibly evidence for this 
phenomenon lurks in the residuals of extant
measurements of the hydrogen spectrum.

Turning to quasar spectra we remark that
\textit{observed} quasar absorption line wavelengths
must first be be corrected by dividing them by the
absorber's redshift factor
$1+z$.  In practice $1+z$ is taken as the ratio of the
observed wavelength of a Bohr transition ($n\neq 0$)
to the corresponding laboratory wavelength.  The
procedure precludes separate determination of the
$\alpha_*$ in the prefactor of  (\ref{new}); any
difference between it and $\alpha$ is absorbed in
the  mentioned  correction, and  $\alpha_*$ and 
$\alpha$ can only be distinguished by the terms
\textit{inside} the square bracket in
Eq.~(\ref{new}).  The correction also makes it
unnecessary to allow for the cosmological evolution
of $m_0$; its variation gets absorbed in the
mentioned  correction.

By substracting the expression  given by
Eq.~(\ref{new}) for
$\lambda_l{}^{-1}$ for  the laboratory line from
$\lambda_z{}^{-1}$,  its analog for the corresponding
redshift--corrected  quasar  line, we get
\begin{equation} {1\over \lambda_z}-{1\over
\lambda_l}={m_0 c\alpha_l{}^3(\alpha_z-\alpha_l)\over
2\pi\hbar}
\cdot\Delta\Big[{1\over n^3(j+{\scriptstyle 1\over
\scriptstyle 2})}-{1\over 2n^4}\Big],
\label{diff}
\end{equation} where we have replaced $\alpha_*$ in
the prefactor by $\alpha_l$ in accordance with our
earlier remarks,  as well as dropped terms of
$\mathcal{O}\big((\alpha_l-\alpha_*)^2\big)$ and
$\mathcal{O}((\alpha_z-\alpha_l)^2$)  in light of the
assumption
$|\chi|\ll 1$.  Contrast Eq.~(\ref{diff}) with the
conventional formula implicit in the Webb group's
analysis,
\begin{equation} {1\over \lambda_z}-{1\over
\lambda_l}={m_0 c\alpha_l{}^3(\alpha_z-\alpha_l)\over
2\pi\hbar}
\cdot\Delta\Big[{1\over n^3(j+{\scriptstyle 1\over
\scriptstyle 2})}-{3\over 4n^4}\Big]
\label{naive}
\end{equation} We observe that formulae (\ref{diff})
and (\ref{naive}) predict exactly the same
$\lambda_z{}^{-1}-
\lambda_l{}^{-1}$ for different members of a
\textit{single}  fine structure multiplet (constant 
$n$ and $\Delta n$) because the formulae differ by a
term ($\propto n^{-4}$) which is constant within the
multiplet.   We conclude that comparison of  fine
structure
\textit{intervals} in a quasar with the 
corresponding ones in the laboratory, with each
multiplet taken separately (as done in older 
work by Webb's group~\cite{webb1} and its predecessors)
cannot distinguish between the $\alpha$ variability from
phase transitions and dynamical
$\alpha$ variability according to the framework.

However, Webb's group has recently analyzed jointly
lines from different multiplets and in different
species~\cite{webb2}.  This procedure
\textit{is} sensitive to the difference between
formulae (\ref{diff}) and (\ref{naive}) and their
nonhydrogen counterparts.  Whereas the  factors in square
brackets in the two formulae  sometimes have similar
values, there are also glaring discrepancies.  Often
the conventional formula underestimates the quantity 
in question by a factor of 2, and for the
transition $2p_{\scriptstyle 3/\scriptstyle 2}
\rightarrow 4s_{\scriptstyle 1/\scriptstyle 2}$  it
underestimates it by a factor of 6.  For the
transition
$2_{\scriptstyle 3/\scriptstyle 2}\rightarrow
3p_{\scriptstyle 1/\scriptstyle 2}$ the conventional
formula \textit{overestimates}
$\lambda_z{}^{-1}- \lambda_l{}^{-1}$ by a factor of
31.5 and gets the sign  wrong.  Clearly if $\alpha$
varies dynamically, a reliable many multiplet
estimate of the cosmological $\alpha$ variation from
quasar absorption spectra must use formula
(\ref{diff}).  And consideration of it is 
indispensable for discriminating between the two
mentioned paradigms of $\alpha$ variability.

I thank Jonathan Oppenheim for useful comments and
the Israel Science Foundation (grant 129/00) for
support.

\end{document}